\documentclass[twocolumn]{aastex63}

\usepackage{upgreek}
\usepackage[T1]{fontenc}

\shorttitle{On the stickiness of CO$_{2}$ and H$_{2}$O ice particles}
\shortauthors{Arakawa \& Krijt}
\graphicspath{{./}{figures/}}

\begin{document}

\title{On the stickiness of CO$_{2}$ and H$_{2}$O ice particles}

\correspondingauthor{Sota Arakawa}
\email{sota.arakawa@nao.ac.jp}

\author[0000-0003-0947-9962]{Sota Arakawa}
\affiliation{Division of Science, National Astronomical Observatory of Japan, Mitaka, Tokyo 181-8588, Japan}

\author[0000-0002-3291-6887]{Sebastiaan Krijt}
\affiliation{College of Engineering, Mathematics and Physics Sciences, University of Exeter, Stocker Rd, Exeter EX4 4QL, UK}



\begin{abstract}

Laboratory experiments revealed that CO$_{2}$ ice particles stick less efficiently than H$_{2}$O ice particles, and there is an order of magnitude difference in the threshold velocity for sticking.
However, the surface energies and elastic moduli of CO$_{2}$ and H$_{2}$O ices are comparable, and the reason why CO$_{2}$ ice particles were poorly sticky compared to H$_{2}$O ice particles was unclear.
Here we investigate the effects of viscoelastic dissipation on the threshold velocity for sticking of ice particles using the viscoelastic contact model derived by Krijt et al.
We find that the threshold velocity for sticking of CO$_{2}$ ice particles reported in experimental studies is comparable to that predicted for perfectly elastic spheres.
In contrast, the threshold velocity for sticking of H$_{2}$O ice particles is an order of magnitude higher than that predicted for perfectly elastic spheres.
Therefore, we conclude that the large difference in stickiness between CO$_{2}$ and H$_{2}$O ice particles would mainly originate from the difference in the strength of viscoelastic dissipation, which is controlled by the viscoelastic relaxation time.

\end{abstract}



\section{Introduction}

Pairwise collisional growth of dust aggregates is the first step of planet formation \citep[e.g.,][]{johansen2014}.
The stickiness and collisional behavior of silicate dust particles/aggregates have been reported in a large number of studies \citep[e.g.,][]{Poppe+2000,Blum+2008,seizinger2013,Kimura+2015,gunkelmann2016,Quadery+2017,planes2020}.
Particles/aggregates composed of H$_{2}$O ice are generally found to be stickier 
\citep[e.g.,][]{Shimaki+2012,Gundlach+2015}, although \citet{Kimura+2020} claimed that H$_{2}$O ice particles might not be stickier than crystalline silicate particles.
This difference in behavior plays an important role in models of dust evolution and planetesimal formation in the inner a few au of circumstellar disks \citep[e.g.,][]{Drazkowska+2017}.

In the cooler outer region of circumstellar disks, not only H$_{2}$O ice but also CO$_{2}$ and/or CO ices are important constituents of icy dust particles \citep[e.g.,][]{oberg2020}.
The condensation temperatures of CO$_{2}$ and CO ices are approximately $70\ {\rm K}$ and $20\ {\rm K}$, respectively \citep[see][]{Okuzumi+2016}.
Using the minimum mass solar nebula model \citep{Hayashi1981}, \citet{Musiolik+2016a} found that the location of the CO$_{2}$ snow line is at $9.3\ {\rm au}$ from the Sun, which is close to the current orbit of Saturn.
\citet{Ali-Dib+2014} also suggests that Uranus and Neptune might be formed near the CO snow line based on the high atmospheric C/H and low N/H ratios.
Therefore, CO$_{2}$ and CO ices may play a crucial role in the planet formation.

In addition, the stickiness of CO$_{2}$ ice particles is of great importance for understanding the dust growth and radial drift behavior in circumstellar disks \citep{Pinilla+2017}.
Recent (sub)millimeter polarimetric observations of circumstellar disks around young stars \citep[e.g.,][]{Kataoka+2017,Stephens+2017} revealed the abundant presence of ${\sim}100\ \upmu{\rm m}$-sized dust particles beyond the H$_{2}$O snow line.
In contrast, the classical theory for dust growth \citep[e.g.,][]{Dominik+1997,Wada+2009} suggests that H$_{2}$O ice particles can grow into significantly larger aggregates when turbulence in a circumstellar disk is moderate.
To solve this discrepancy, \citet{Okuzumi+2019} proposed an idea that the low stickiness of CO$_{2}$ ice particles reported by \citet{Musiolik+2016a,Musiolik+2016b} might be the key to explain the small size of dust particles observed in circumstellar disks.

Laboratory experiments by \citet{Musiolik+2016a,Musiolik+2016b} revealed that CO$_{2}$ ice particles are less sticky compared to H$_{2}$O ice particles.
\citet{Pinilla+2017} and \citet{Okuzumi+2019} proposed that the large difference in stickiness between H$_{2}$O and CO$_{2}$ ice particles would originate from the difference in the dipole moment.
In other words, the low threshold velocity for sticking of CO$_{2}$ ice particles is due to the small surface free energy of {\it apolar} CO$_{2}$ ice.
However, we note that the literature value of the surface free energy of CO$_{2}$ ice \citep[$80\ {\rm mJ}\ {\rm m}^{-2}$;][]{Wood1999} is comparable to that of H$_{2}$O ice \citep[$100\ {\rm mJ}\ {\rm m}^{-2}$;][]{Israelachvili2011}.
In addition, the values of elastic properties (i.e., the Young's modulus and Poisson ratio) are also similar between two materials (see Section \ref{sec.3}).
In the framework of \citet{Dominik+1997}, one would then expect the threshold velocity for sticking to be similar for H$_{2}$O and CO$_{2}$ ices.

In this study, we investigate another possibility to explain the low threshold velocity for sticking of CO$_{2}$ ice particles compared to that of H$_{2}$O ice particles.
\citet{Krijt+2013} constructed a viscoelastic contact model, which is the advanced version of the contact theory for perfectly elastic spheres \citep[e.g.,][]{Johnson+1971,Wada+2007}.
The viscoelastic contact model of \citet{Krijt+2013} takes into account a crack propagation at the edge of the contact and an energy dissipation arising from viscoelastic behavior beneath the contact.
Applying this model to water ice particles, \citet{Gundlach+2015} found that the threshold velocity for sticking is up to an order of magnitude higher than that predicted from the theory for perfectly elastic spheres.
Therefore, we can potentially explain the large difference in stickiness between H$_{2}$O and CO$_{2}$ ice particles reported by \citet{Musiolik+2016a,Musiolik+2016b} if CO$_{2}$ ice particles follow more closely the contact theory for perfectly elastic adhesive spheres.

The structure of this paper is as follows.
In Section \ref{sec.2}, we review the viscoelastic contact model derived by \citet{Krijt+2013}.
In Section \ref{sec.3}, we summarize the material properties of CO$_{2}$ ice.
In Section \ref{sec.4}, we show the typical results for collisions between two viscoelastic spheres.
In Section \ref{sec.5}, we calculate the threshold velocity for sticking and compare our numerical results with experimental data reported by \citet{Musiolik+2016a,Musiolik+2016b}.
In Section \ref{sec.6}, we evaluate the critical velocity for collisional growth/fragmentation of dust aggregates.
Implications of our results are discussed in Section \ref{sec.7}, and we conclude in Section \ref{sec.8}.

\section{Contact Model}
\label{sec.2}

The contact model used in this study is identical to what \citet{Krijt+2013} derived.
In Section \ref{sec.2}, we briefly summarize their viscoelastic contact model.

\subsection{Elastic strain energy stored in a contact}

When two elastic spheres are pressed together, they will deform locally and share a circular contact area with radius, $a$.
The pressure distribution in the contact area, $p {\left( r \right)}$, is given as a function of the distance from the center of the contact, $r$, as follows \citep{Muller+1980}:
\begin{equation}
p {\left( r \right)} = \frac{E^{*}}{\pi R} \frac{a^{2} - 2 r^{2} + R \delta}{\sqrt{a^{2} - r^{2}}},
\label{eq.p}
\end{equation}
where $\delta$ is the mutual approach, $R$ is the reduced particle radius, and $E^{*}$ is the elastic contact modulus.
For a contact between two spheres with the same radius and material, $R$ and $E^{*}$ are given by $R = R_{1} / 2$ and $E^{*} = E / {\left[ 2 {\left( 1 - \nu^{2} \right)} \right]}$, where $R_{1}$ is the particle radius, $E$ is the (relaxed) Young's modulus, and $\nu$ is the Poisson ratio. 
Then the elastic strain energy stored in the contact, $U_{\rm E}$, is given by \citep{Muller+1980}
\begin{eqnarray}
U_{\rm E} & = & \frac{1}{2} \int_{0}^{a} {\rm d}r\ 2 \pi r {p {\left( r \right)}} {w {\left( r \right)}} \nonumber \\
          & = & \frac{E^{*} a^{3}}{3 R} {\left[ \delta {\left( \frac{3 \delta R}{a^{2}} - 1\right)} - \frac{a^{2}}{5 R} {\left( \frac{5 \delta R}{a^{2}} - 3 \right)} \right]},
\end{eqnarray}
where $w {\left( r \right)} = \delta - {r^{2}} / {\left( 2 R \right)}$ is the deformation of the surface of spheres.

\subsection{Johnson--Kendall--Roberts theory}

\citet{Johnson+1971} introduced a surface energy term, $U_{\rm S}$, to describe a contact between adhesive particles:
\begin{equation}
U_{\rm S} = - \pi a^{2} \gamma.
\end{equation}
Assuming that the contact area changes quasistatically, ${\partial U_{\rm S}} / {\partial a}$ is given by
\begin{equation}
\frac{\partial U_{\rm S}}{\partial a} = - 2 \pi a \gamma.
\label{eq.gamma}
\end{equation}
It is known that an equilibrium exists in the framework of Johnson--Kendall--Roberts theory \citep[hereinafter referred to as JKR theory;][]{Johnson+1971}.
If there are no external forces, the contact radius at the equilibrium is
\begin{equation}
a_{\rm eq} = {\left( \frac{9 \pi \gamma R^{2}}{2 E^{*}} \right)}^{1/3},
\end{equation}
and the mutual approach at the equilibrium is given by
\begin{eqnarray}
\delta_{\rm eq} & = & \frac{{a_{\rm eq}}^{2}}{R} - \sqrt{\frac{2 \pi \gamma a_{\rm eq}}{E^{*}}} \nonumber \\
                & = & {\left( \frac{3 \pi^{2} R \gamma^{2}}{4 {E^{*}}^{2}} \right)}^{1/3}.
\end{eqnarray}

\citet{Johnson+1971} assumed that there are no forces acting outside the contact area for simplicity.
This treatment works well when the Tabor parameter, $\mu$, is sufficiently large, i.e., $\mu \gg 1$ \citep{Tabor1977}.
The Tabor parameter is defined as
\begin{equation}
\mu = \frac{1}{\epsilon} {\left( \frac{R \gamma^{2}}{{E^{*}}^{2}} \right)}^{1/3},
\end{equation}
where $\epsilon$ is the range of action of the surface forces and $\gamma$ is the surface energy.
We set $\epsilon = 0.3\ {\rm nm}$ in this study \citep{Krijt+2013}.
Using material properties of CO$_2$ ice, we found that $\mu \simeq 9.5$ for $R_{1} = 60\ \upmu{\rm m}$, and JKR theory could be appliable for (sub)micron-sized CO$_2$ ice particles.

\subsection{Viscoelastic crack velocity}

If spheres are made of perfectly elastic materials, we can use the surface energy term introduced by \citet{Johnson+1971}.
However, when the material is viscoelastic, the propagating cracks have non-zero velocities and we can no longer use Equation (\ref{eq.gamma}) to calculate the surface energy term.
In this case, the energy released/absorbed when the crack is closed/opened, ${\partial U_{\rm S}^{*}} / {\partial a}$, is
\begin{equation}
\frac{\partial U_{\rm S}^{*}}{\partial a} = - 2 \pi a G_{\rm eff},
\end{equation}
where $G_{\rm eff}$ is the {\it apparent} surface energy for bonding/cracking.
The apparent surface energy, $G_{\rm eff}$, depends on the crack velocity, $\dot{a}$.
Here we introduce the normalized apparent surface energy, $\beta$, and the normalized crack velocity, $v$:
\begin{eqnarray}
\beta & \equiv & \frac{G_{\rm eff}}{\gamma}, \\
v     & \equiv & \frac{{\sigma_{0}}^{2} T_{\rm vis}}{E^{*} \gamma} \dot{a},
\end{eqnarray}
where $\sigma_{0} = \gamma / \epsilon$ is the attractive force acting across the crack and $T_{\rm vis}$ is the viscoelastic relaxation time \citep{Greenwood2004,Krijt+2013}.
The normalized apparent surface energy, $\beta$, is a function of $k$ and $v$:
\begin{equation}
\beta = \beta {\left( k, v \right)},
\end{equation}
where $k$ is the ratio of relaxed to instantaneous elastic moduli.\footnote{
The elastic strain distribution of viscoelastic media due to the instant application of loads should be given by the {\it instantaneous} elastic modulus.
In contrast, even if the loads remain constant, the strain distribution grows according to the creep, and the final strain distribution is given by the {\it relaxed} elastic modulus.
Therefore we distinguish these two elastic moduli.
The fact that the instantaneous modulus is much larger than the relaxed modulus (i.e., $k \ll 1$) means that the stress relaxation is much faster than creep \citep[see][and references therein]{Baney+1999,Greenwood2004}.
}
In practice, $k \ll 1$ and we set $k = 0.01$ as a fiducial value \citep{Greenwood2004}.
We note that \citet{Krijt+2013} set $k = 0.02$ instead; however, this small difference in $k$ hardly changes our numerical results (see Figure \ref{fig4}).
The normalized crack velocity is also a function of $k$ and $\beta$:
\begin{equation}
v = v {\left( k, \beta \right)}.
\label{eq.v}
\end{equation}
Figure \ref{fig1} shows the dependence of $v$ on $\beta$ for different values of $k$ \citep[see][]{Greenwood2004}.
We describe how did we calculate $v {\left( k, \beta \right)}$ in Appendix \ref{App.A}.

\subsection{Apparent surface energy in equilibrium}

\citet{Krijt+2013} assumed that the crack velocity, $\dot{a}$, and the apparent surface energy, $G_{\rm eff}$, adjust themselves to satisfy the equilibrium contact condition:
\begin{equation}
\frac{\partial}{\partial a} {\left( U_{\rm E} + U_{\rm S}^{*} \right)} = 0,
\end{equation}
which can be solved to give
\begin{equation}
G_{\rm eff} = \frac{E^{*}}{2 \pi a R^{2}} {\left( a^{2} - \delta R \right)}^{2}.
\label{eq.Geff}
\end{equation}
Then the apparent surface energy, $G_{\rm eff} = G_{\rm eff} {\left( a, \delta \right)}$, is given by Equation (\ref{eq.Geff}) and the crack velocity, $\dot{a} = \dot{a} {\left( a, \delta \right)}$, is given by Equation (\ref{eq.v}).

\subsection{Elastic and dissipative forces}

The pressure distribution in the contact area is given by Equation (\ref{eq.p}), and the integral over the contact area yields the elastic force between two particles, $F_{\rm E}$:
\begin{eqnarray}
F_{\rm E} & = & \int_{0}^{a} {\rm d}r\ 2 \pi r {p {\left( r \right)}} \nonumber \\
          & = & \frac{2 E^{*}}{3 R} {\left( 3 a \delta R - a^{3} \right)}.
\end{eqnarray}
When two viscoelastic particles collide and deform, a significant amount of energy could be dissipated.
Following \citet{Krijt+2013}, we write the dissipative force as follows:
\begin{eqnarray}
F_{\rm D} & = & \int_{0}^{a} {\rm d}r\ 2 \pi r A \frac{\partial {p {\left( r \right)}}}{\partial \delta} \dot{\delta} \nonumber \\
          & = & 2 A E^{*} a \dot{\delta},
\end{eqnarray}
where $A = T_{\rm vis} / \nu^{2}$ \citep{Brilliantov+1996,Brilliantov+2007}.
The dissipative force, $F_{\rm D}$, depends on $a$ and $\dot{\delta}$, and it acts as a drag term.

For a head-on collision of identical spheres, the time evolution of the mutual approach is given by
\begin{equation}\label{eq.d_dt}
\ddot{\delta} = - \frac{1}{m^{*}} {\left( F_{\rm E} + F_{\rm D} \right)},
\end{equation}
where $m^{*}$ is the reduced mass, which is $m^{*} = m_{1} / 2$ for a contact between two spheres with the same radius and material.
The mass of each particle is given by $m_{1} = 4 \pi \rho {R_{1}}^{3} / 3$, where $\rho = 1560\ {\rm kg}\ {\rm m}^{-3}$ is the material density of CO$_{2}$ ice \citep{Mazzoldi+2008}.

Ignoring long-range forces, the moment of first contact is taken as $t = 0$, and the initial condition for the mutual approach is $\delta = 0$ and $\dot{\delta} = V_{\rm in}$, where $V_{\rm in}$ is the collision velocity.
The normalized crack velocity, $v = v {\left( k, \beta \right)}$, is defined within the range of $k < \beta < 1 / k$ (see Appendix \ref{App.A}) and the apparent surface energy satisfies $G_{\rm eff} = \beta \gamma > 0$.
Equation (\ref{eq.Geff}) is rewritten as $G_{\rm eff} = E^{*} a^{3} / {\left( 2 \pi R^{2} \right)}$ when $\delta = 0$, and it does not allow $a = 0$ as the initial condition.
For our numerical integrations, we set $\beta = {\left( 1 + \varepsilon \right)} k$ at $t = 0$ as \citet{Krijt+2013} assumed.
We use the small value of $\varepsilon = 0.01$.
We end integrations when $\beta = {\left( 1 - \varepsilon \right)} / k$, and two spheres will separate immediately.
These assumptions are justified as the contact evolves rapidly near $a \simeq 0$ and $\delta$ hardly changes \citep[see Appendix B of][]{Krijt+2013}.

\section{Material Properties of Carbon Dioxide Ice}
\label{sec.3}

In Section \ref{sec.3}, we discuss the elastic and adhesive material properties of CO$_{2}$ and H$_{2}$O ices, needed for solving Equation (\ref{eq.d_dt}).
Following \citet{Gundlach+2015}, we treat the main viscoelastic parameter $T_{\rm vis}$ as a free parameter that may depend on particle size.

\subsection{Surface free energy}

The surface free energies of crystals are proportional to their sublimation energies \citep[e.g.,][]{Shuttleworth1949,Benson+1964}.
Using the crystal structure and the latent heat of sublimation of CO$_{2}$ ice, \citet{Wood1999} theoretically estimated the surface free energy of CO$_{2}$ ice as $\gamma_{\rm SV} = 80\ {\rm mJ}\ {\rm m}^{-2}$.
This value is widely used in the studies of CO$_{2}$ clouds in the martian atmosphere \citep[e.g.,][]{Maattanen+2005,Nachbar+2016,Mangan+2017}.
\citet{Glandorf+2002} also obtained an approximate value of $\gamma_{\rm SV}$ by using the Antonoff's rule, and the surface free energy is estimated as $67\ {\rm mJ}\ {\rm m}^{-2}$, which is in reasonable agreement with that obtained by \citet{Wood1999}.\footnote{
\citet{Wood1999} also tested the validity of the technique for estimating surface energy.
Using the technique, they obtained that the surface energy of H$_{2}$O ice is $128\ {\rm mJ}\ {\rm m}^{-2}$ for the prism face (and $120\ {\rm mJ}\ {\rm m}^{-2}$ for the basal face).
This theoretical estimate shows good agreement with the canonical value obtained from experiments \citep[i.e., $100\ {\rm mJ}\ {\rm m}^{-2}$;][]{Israelachvili2011}.
}

For a contact between two spheres made of same material, the surface energy, $\gamma$, is (approximately) twice the surface free energy \citep{Johnson+1971}:
\begin{equation}
\gamma = 2 \gamma_{\rm SV}.
\end{equation}

For H$_{2}$O ice, the canonical value of $\gamma_{\rm SV}$ is $\gamma_{\rm SV} = 100\ {\rm mJ}\ {\rm m}^{-2}$ \citep{Israelachvili2011}.
We note, however, that the surface free energy of H$_{2}$O ice is still under debate (see Section \ref{sec.5.2}).

\subsection{Young's modulus and Poisson ratio}

As the longitudinal and transversal velocities of sound, $v_{\rm lg}$ and $v_{\rm ts}$, are related to the elastic properties, we can calculate the Young's modulus and Poisson ratio from the results of sound velocity measurements.
These sound velocities, $v_{\rm lg}$ and $v_{\rm ts}$, are given by \citep[e.g.,][]{Han+2004}
\begin{eqnarray}
v_{\rm lg} & = & \sqrt{\frac{1}{\rho} {\left( \mathcal{K} + \frac{4}{3} \mathcal{G} \right)}}, \\
v_{\rm ts} & = & \sqrt{\frac{\mathcal{G}}{\rho}},
\end{eqnarray}
where $\mathcal{K}$ is the bulk modulus and $\mathcal{G}$ is the shear modulus.
Both $\mathcal{K}$ and $\mathcal{G}$ can be rewritten by using $E$ and $\nu$ as follows:
\begin{eqnarray}
\mathcal{K} & = & \frac{E}{3 {\left( 1 - 2 \nu \right)}}, \\
\mathcal{G} & = & \frac{E}{2 {\left( 1 + \nu \right)}}.
\end{eqnarray}
\citet{Yamashita+1997} measured the longitudinal and transversal velocities of sound in CO$_{2}$ ice and obtained $v_{\rm lg} = 2900\ {\rm m}\ {\rm s}^{-1}$ and $v_{\rm ts} = 1650\ {\rm m}\ {\rm s}^{-1}$ at the temperature of $80\ {\rm K}$ \citep[see][]{Musiolik+2016a}.
Then the Young's modulus and Poisson ratio are $E = 10.7\ {\rm GPa}$ and $\nu = 0.26$, respectively.

For H$_{2}$O ice, the literature values of $E = 7\ {\rm GPa}$ and $\nu = 0.25$ are widely used in previous studies \citep[e.g.,][]{Dominik+1997,Wada+2007,Gundlach+2015}.

\section{Sticking, Bouncing, and Double Collisions}
\label{sec.4}

The contact model reviewed in Section \ref{sec.2} can be used to calculate the time evolution of the contact between two colliding spheres.
In Section \ref{sec.4}, we show the typical results for collisions between two equal-sized spheres of CO$_{2}$ ice.

We begin by setting $R_{1} = 60\ \upmu{\rm m}$ and $T_{\rm vis} = 10^{-9}\ {\rm s}$, and exploring a range of impact collision velocities $V_{\rm in}$.
We found that there are three types of collision outcomes, namely, sticking collisions, bouncing collisions, and double collisions.
Similar results are also reported in Sections 3.1 and 3.2 of \citet{Krijt+2013}.

\subsection{Sticking collision}

\begin{figure*}
\plottwo{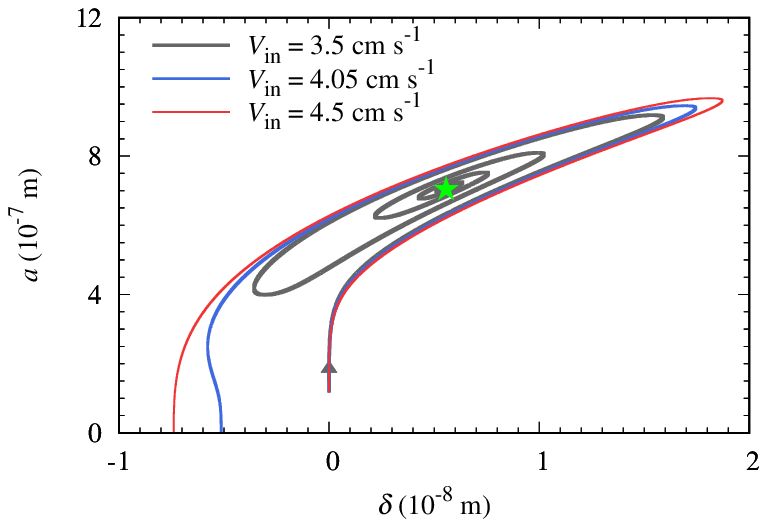}{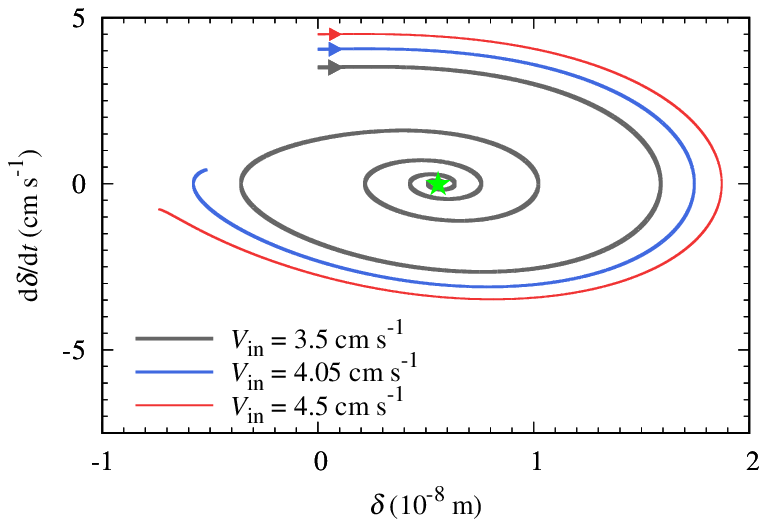}
\caption{
Time evolution of the contact radius, $a$, the mutual approach, $\delta$, and the approaching velocity, $\dot{\delta}$, for head-on collisions.
The green stars mark the equilibrium point in JKR theory ($a = a_{\rm eq}$, $\delta = \delta_{\rm eq}$, and $\dot{\delta} = 0$).
The left panel shows the evolution in $\delta$--$a$ plane, and the right panel is the evolution in $\delta$--$\dot{\delta}$ plane, respectively.
The grey lines represent the evolutionary track for the case of $V_{\rm in} = 3.5\ {\rm cm}\ {\rm s}^{-1}$, resulting in sticking.
The blue lines are for the case of $V_{\rm in} = 4.05\ {\rm cm}\ {\rm s}^{-1}$, resulting in double collision in the collisional sequence.
The red lines are for the case of $V_{\rm in} = 4.5\ {\rm cm}\ {\rm s}^{-1}$, resulting in bouncing.
}
\label{fig2}
\end{figure*}

The grey lines of Figure \ref{fig2} show the evolution of the contact radius, $a$, the mutual approach, $\delta$, and the approaching velocity, $\dot{\delta}$, for a head-on collision at $V_{\rm in} = 3.5\ {\rm cm}\ {\rm s}^{-1}$.
The green stars mark the equilibrium point in JKR theory ($a = a_{\rm eq}$, $\delta = \delta_{\rm eq}$, and $\dot{\delta} = 0$).
At the start of the collision, $t = 0$, the mutual approach is $\delta = 0$ and the contact radius is given by
\begin{equation}
a = {\left[ \frac{2 \pi R^{2}}{E^{*}} {\left( 1 + \varepsilon \right)} k \gamma \right]}^{1/3}.
\end{equation}
The contact radius initially grows very rapidly, as $a$ increases to $a \simeq 0.3\ \upmu{\rm m}$ with $\delta$ hardly changing.
\citet{Krijt+2013} described the details of the behavior of the viscoelastic contact, by comparing with that of JKR theory \citep[e.g.,][]{Wada+2007}.

The most important difference between our viscoelastic contact model and JKR theory is whether the kinetic energy dissipates during contact or not.
For the case of $V_{\rm in} = 3.5\ {\rm cm}\ {\rm s}^{-1}$, the spheres cannot separate and instead oscillate back and forth.
In $\delta$--$a$ and $\delta$--$\dot{\delta}$ planes, the contact spirals toward the equilibrium point of JKR theory due to the dissipative effects when we use the viscoelastic contact model.
In the framework of JKR theory, in contrast, the oscillation would not be dampened.
The dissipative effects increase the threshold velocity for sticking, $V_{\rm stick}$ (see Section \ref{sec.5}).

\subsection{Bouncing collision}

Even if the dissipative effects work, collisions of two spheres will result in bouncing as the collision velocity is increased.
The red lines of Figure \ref{fig2} show the evolution of $a$, $\delta$, and $\dot{\delta}$, for a head-on collision at $V_{\rm in} = 4.5\ {\rm cm}\ {\rm s}^{-1}$.
In this case, the contact radius finally becomes $a \simeq 0$, and the mutual approach and the approaching velocity are $\delta > 0$ and $\dot{\delta} < 0$ at the end of the contact.
At that point, the spheres separate and move away from each other at a velocity $V_\mathrm{out}$ (see Section \ref{sec.4.4}).

\subsection{Double collision}

There exists a narrow range of impact velocities for which we observe a ``double collision''.
This double collision occurs as a result of energy dissipations and viscoelastic cracking.
The blue lines of Figure \ref{fig2} show the evolution of $a$, $\delta$, and $\dot{\delta}$, for a head-on collision at $V_{\rm in} = 4.05\ {\rm cm}\ {\rm s}^{-1}$.
In this case, the mutual approach and approaching velocity are $\delta > 0$ and $\dot{\delta} > 0$ at the end of the contact.
As $\dot{\delta} > 0$, two spheres are expected to recollide after their separation.
We therefore named this outcome as the ``double collision''.
We note that the collision velocity of the second collision is much lower than that of the first collision because of dissipative effects, and the second collision should result in sticking.

\subsection{Coefficient of restitution}
\label{sec.4.4}

We use the coefficient of restitution, $e$, to describe colllision outcomes.
The definition of $e$ is
\begin{equation}
e \equiv - \frac{V_{\rm out}}{V_{\rm in}},
\end{equation}
where $V_{\rm out}$ is the approaching velocity at the end of the contact.
We set $e = 0$ for sticking collisions.
For double collisions, negative values of the coefficient of restitution will be obtained from numerical calculations.
We note, however, that the second collision may occur immediately after the first collision and the final outcome of the collisional sequence is sticking.
Then we can imagine that the ``observed'' value of the coefficient of restitution in laboratory experiments is $e = 0$ for double collisions.

\begin{figure}
\plotone{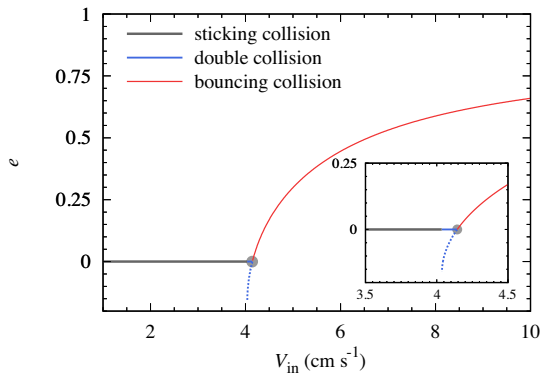}
\caption{
Variations of the coefficient of restitution, $e$, with the collision velocity, $V_{\rm in}$, for head-on collisions of CO$_{2}$ ice particles.
Particle radius of $R_{1} = 60\ \upmu{\rm m}$ and a relaxation time of $T_{\rm vis} = 10^{-9}\ {\rm s}$ are assumed here.
The grey, blue, and red solid lines represent the values of $e$ which are expected to be observed as final outcomes, and the blue dashed line is the value of $e$ obtained from numerical calculations (for double collisions).
The grey point indicates the threshold velocity for sticking, $V_{\rm stick} = 4.14\ {\rm cm}\ {\rm s}^{-1}$.
}
\label{fig3}
\end{figure}

Figure \ref{fig3} shows the variations of $e$ with $V_{\rm in}$ for $R_{1} = 60\ \upmu{\rm m}$ and $T_{\rm vis} = 10^{-9}\ {\rm s}$, a transition from sticking collisions to double collisions occurs at $V_{\rm in} = 4.04\ {\rm cm}\ {\rm s}^{-1}$, and a transition from double collisions to bouncing collisions occurs at $V_{\rm in} = 4.14\ {\rm cm}\ {\rm s}^{-1}$.
In this case, we obtain the threshold velocity for sticking as $V_{\rm stick} = 4.14\ {\rm cm}\ {\rm s}^{-1}$.
As $V_{\rm stick}$ depends on the particle radius and material properties including $\gamma_{\rm SV}$ and $T_{\rm vis}$, we can estimate the relaxation time of viscoelastic particles from literature values of $V_{\rm stick}$ which are experimentally determined \citep[e.g.,][]{Krijt+2013,Gundlach+2015}.

\section{Threshold Velocity for Sticking}
\label{sec.5}

In section \ref{sec.5}, we calculate the threshold velocity for sticking using the viscoelastic contact model, and we also compare our numerical results with experimental data reported by \citet{Musiolik+2016a,Musiolik+2016b}.
We show that $V_{\rm stick}$ of both CO$_{2}$ and H$_{2}$O ice particles observed in experiments are consistent with the theoretical prediction from the viscoelastic contact model.
Especially, $V_{\rm stick}$ of H$_{2}$O ice particles can be reproduced only when we consider the dissipative effects.

\subsection{Carbon dioxide ice}
\label{sec.5.1}

\citet{Musiolik+2016a} performed laboratory experiments of collisions of CO$_{2}$ ice particles within a vacuum chamber at a temperature of $80\ {\rm K}$.
The collision velocities are below $2.5\ {\rm m}\ {\rm s}^{-1}$, and the typical radius of the particles is $R_{1} = 60\ \upmu{\rm m}$ when we focus on the collisions of small particles whose radii are less than $150\ \upmu{\rm m}$.\footnote{
The size distribution of the CO$_{2}$ ice particles is shown in Figure 4 of \citet{Musiolik+2016a}, and 80\% of all particles are within the size range of $40\ \upmu{\rm m} < R_{1} < 120\ \upmu{\rm m}$.
}
They found that the threshold velocity for sticking is $V_{\rm stick} = {\left( 0.04 \pm 0.02 \right)}\ {\rm m}\ {\rm s}^{-1}$, although the uncertainty is large.

\begin{figure}
\plotone{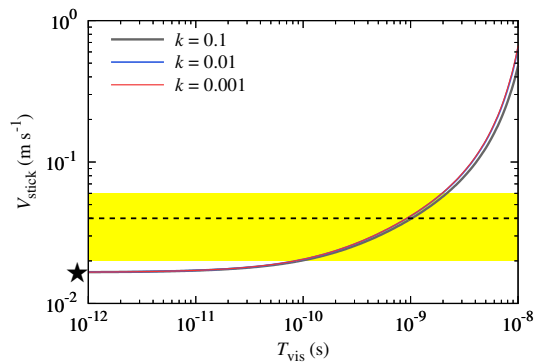}
\caption{
Dependence of the threshold velocity for sticking, $V_{\rm stick}$, on the relaxation time, $T_{\rm vis}$, for different values of $k$.
The black dashed line represents the threshold velocity for sticking obtained from laboratory experiments and the yellow shaded region shows the uncertainty: $V_{\rm stick} = {\left( 0.04 \pm 0.02 \right)}\ {\rm m}\ {\rm s}^{-1}$ \citep[see][]{Musiolik+2016a}.
The black star indicates the threshold velocity for sticking inferred from JKR theory, $V_{\rm stick,\, JKR}$ \citep[e.g.,][]{Thornton+1998,Wada+2007}.
The typical radius of CO$_{2}$ ice particles used in \citet{Musiolik+2016a} is $R_{1} = 60\ \upmu{\rm m}$.
}
\label{fig4}
\end{figure}

Figure \ref{fig4} shows the dependence of $V_{\rm stick}$ on $T_{\rm vis}$ for different values of $k$.
As mentioned in \citet{Greenwood2004} and \citet{Krijt+2013}, the evolution of contact radius is almost independent of $k$ except for the start and end of the contact.
Then the collision outcomes hardly depend on the choice of $k$ as long as we set $k \ll 1$ (see Appendix \ref{App.A}).

As shown in Figure \ref{fig4}, $V_{\rm stick}$ hardly changes when $T_{\rm vis} \lesssim 10^{-11}\ {\rm s}$.
In this case, $V_{\rm stick}$ is almost identical to that of JKR theory.
According to \citet{Thornton+1998}, in the framework of JKR theory, the threshold velocity for sticking is given by
\begin{eqnarray}
V_{\rm stick,\, JKR} & = & {\left( \frac{14.18}{m^{*}} \right)}^{1/2} {\left( \frac{\gamma^{5} R^{4}}{{E^{*}}^{2}} \right)}^{1/6} \nonumber \\
                     & = & 1.66 \times 10^{-2}\ {\left( \frac{R_{1}}{60\ \upmu{\rm m}} \right)}^{- 5/6}\ {\rm m}\ {\rm s}^{-1}.
\end{eqnarray}
The black star plotted in Figure \ref{fig4} indicates the value of $V_{\rm stick,\, JKR}$, and it is clear that $V_{\rm stick} \to V_{\rm stick,\, JKR}$ for the short-$T_{\rm vis}$ limit, $T_{\rm vis} \to 0$.

In contrast, when $T_{\rm vis} \gtrsim 10^{-9}\ {\rm s}$, the threshold velocity for sticking is several times higher than that predicted from JKR theory.
The increase of $V_{\rm stick}$ with increasing of $T_{\rm vis}$ is also reported in previous studies \citep{Krijt+2013,Gundlach+2015}, and our results shown in Figure \ref{fig4} are consistent with their results.
Assuming that $k = 0.01$, we can obtain the suitable range of $T_{\rm vis}$ to reproduce $V_{\rm stick}$ reported by \citet{Musiolik+2016a} as follows:
\begin{equation}
8.5 \times 10^{-11}\ {\rm s} \leq T_{\rm vis} \leq 1.97 \times 10^{-9}\ {\rm s};
\label{T_vis_CO2}
\end{equation}
though we do not reject the possibility that $T_{\rm vis} \ll 10^{-10}\ {\rm s}$ and $V_{\rm stick}$ is nearly identical to $V_{\rm stick,\, JKR}$.

In numerical calculations, we assumed that CO$_{2}$ ice particles are spherical and the viscoelastic contact theory for spheres is appliable. 
We acknowledge, however, that CO$_{2}$ ice particles used in \citet{Musiolik+2016a} are not spherical.
Although \citet{Blum+2000} suggested that irregular grains are slightly stickier than spherical grains, \citet{Musiolik+2016a} mentioned that the effect of the irregular shape may be negligible.

\citet{Musiolik+2016a,Musiolik+2016b} did not report the surface roughness of ice grains, however, it might alter the threshold velocity for sticking \citep[e.g.,][]{Nagaashi+2018}.
Although our results for both CO$_{2}$ and H$_{2}$O ice particles are consistent with the cases for smooth particles (see Sections \ref{sec.5.2} and \ref{sec.5.3}), we need to assess the effect of the surface roughness in future.

It should also be noted that whether CO$_{2}$ ice particles were monolithic grains or aggregates is unknown.
In this study, however, we assume that CO$_{2}$ ice particles whose radii are $R_{1} \simeq 60\ \upmu{\rm m}$ are monolithic.
This is because the critical velocity for collisional growth/fragmentation, $V_{\rm frag}$, should be several times higher than $0.04\ {\rm m}\ {\rm s}^{-1}$ when CO$_{2}$ ice particles are aggregates, even if the monomer grains of these aggregates behaved as perfectly elastic spheres (see Section \ref{sec.7.1} for details).

\subsection{Water ice}
\label{sec.5.2}

\citet{Musiolik+2016b} also performed laboratory experiments of collisions of pure H$_{2}$O ice particles (and mixtured ice particles of H$_{2}$O--CO$_{2}$) within a vacuum chamber at a temperature of $80\ {\rm K}$.
The typical radius of the particles is $R_{1} = 90\ \upmu{\rm m}$, and their experimental results suggest that $V_{\rm stick} \sim 0.73\ {\rm m}\ {\rm s}^{-1}$ for pure H$_{2}$O ice particles.

Material properties of H$_{2}$O ice are reported in a large number of previous studies.
We set $E = 7\ {\rm GPa}$, $\nu = 0.25$, and $\rho = 930\ {\rm kg}\ {\rm m}^{-3}$ \citep{Gundlach+2015}.
The surface free energy of H$_{2}$O ice is still under debate.
The canonical value of $\gamma_{\rm SV}$ used in numerical simulations \citep[e.g.,][]{Wada+2013,Sirono+2017,Tatsuuma+2019} is $\gamma_{\rm SV} = 100\ {\rm mJ}\ {\rm m}^{-2}$ \citep{Israelachvili2011}.
Measurements of the critical rolling friction force of $\upmu{\rm m}$-sized H$_{2}$O ice particles suggest that $\gamma_{\rm SV} = 190\ {\rm mJ}\ {\rm m}^{-2}$ \citep{Gundlach+2011}, although the value of $\gamma_{\rm SV}$ depends on the assumed value of the critical rolling displacement \citep[see][]{Krijt+2014}.
Moreover, tensile strength measurements in a low-temperature environment \citep{Gundlach+2018} suggest that $\gamma_{\rm SV} = 20\ {\rm mJ}\ {\rm m}^{-2}$ at low temperatures below $150\ {\rm K}$.
\citet{musiolik2019} also reported that $\gamma_{\rm SV}$ at $175\ {\rm K}$ is one to two orders of magnitude lower than the canonical value based on their pull-off measurements of mm-sized water ice grains.
Then we parameterize $\gamma_{\rm SV}$ in our calculations.

\begin{figure}
\plotone{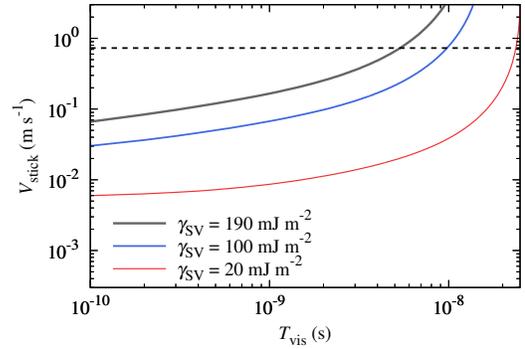}
\caption{
Dependence of the threshold velocity for sticking, $V_{\rm stick}$, on the relaxation time, $T_{\rm vis}$, for different values of $\gamma_{\rm SV}$.
The black dashed line represents the threshold velocity for sticking obtained from laboratory experiments.
The typical radius of H$_{2}$O ice particles used in \citet{Musiolik+2016b} is $R_{1} = 90\ \upmu{\rm m}$.}
\label{fig5}
\end{figure}

Figure \ref{fig5} then shows the dependence of $V_{\rm stick}$ on $T_{\rm vis}$ for different values of $\gamma_{\rm SV}$.
We found that we cannot explain the reported value of $V_{\rm stick}$ by using JKR theory, that is, the contact model for perfectly elastic adhesive spheres.
Assuming that the range of the surface free energy is $20\ {\rm mJ}\ {\rm m}^{-2} \le \gamma_{\rm SV} \le 190\ {\rm mJ}\ {\rm m}^{-2}$, the required value of $T_{\rm vis}$ is
\begin{equation}
5.3 \times 10^{-9}\ {\rm s} \leq T_{\rm vis} \leq 2.37 \times 10^{-8}\ {\rm s},
\label{T_vis_H2O}
\end{equation}
and $T_{\rm vis}$ of H$_{2}$O ice particles with $R_{1} = 90\ \upmu{\rm m}$ may be an order of magnitude larger than that of CO$_{2}$ ice particles with $R_{1} = 60\ \upmu{\rm m}$.

For H$_{2}$O ice particles with $R_{1} = 1.5\ \upmu{\rm m}$, \citet{Gundlach+2015} revealed that $T_{\rm vis} = 1 \times 10^{-10}\ {\rm s}$ is plausible to reproduce the value of $V_{\rm stick} = {\left( 9.6 \pm 0.3 \right)}\ {\rm m}\ {\rm s}^{-1}$ obtained from their experiments.
In Section \ref{sec.5.3}, we discuss the dependence of $T_{\rm vis}$ on $R_{1}$.

\subsection{Relaxation time}
\label{sec.5.3}

The relaxation time is a fitted parameter in this study because we do not know how $T_{\rm vis}$ relates to other fundamental material properties.
We note, however, that there is an empirical relation between $T_{\rm vis}$ and $R_{1}$ \citep{Krijt+2013,Gundlach+2015}.
\citet{Gundlach+2015} reported that the relaxation times obtained by \citet{Krijt+2013} is consistent with a relation between $T_{\rm vis}$ and $R_{1}$, that is, $T_{\rm vis} \propto {R_{1}}^{1.11}$.

For H$_{2}$O ice particles with $R_{1} = 1.5\ \upmu{\rm m}$, \citet{Gundlach+2015} revealed that $T_{\rm vis} = 1 \times 10^{-10}\ {\rm s}$.
Therefore, the size-dependent relaxation time of H$_{2}$O ice particles may be given by \citep{Gundlach+2015}
\begin{equation}
T_{\rm vis,\, H_{2}O} = 1 \times 10^{-10}\ {\left( \frac{R_{1}}{1.5\ \upmu{\rm m}} \right)}^{1.11}\ {\rm s},
\label{T_vis_R1_H2O}
\end{equation}
and this equation yields $T_{\rm vis,\, H_{2}O} = 9.4 \times 10^{-9}\ {\rm s}$ for H$_{2}$O ice particles with $R_{1} = 90\ \upmu{\rm m}$.
This relation shows excellent agreement with our numerical results (see Relation \ref{T_vis_H2O}).

\begin{figure}
\plotone{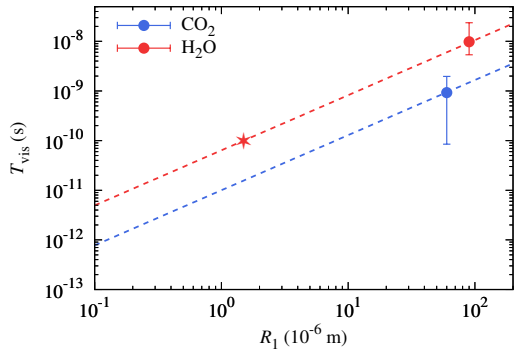}
\caption{
Dependence of the relaxation time, $T_{\rm vis}$, on the particle radius, $R_{1}$, for both CO$_{2}$ and H$_{2}$O ice particles.
The blue and red points with error bars are the calculated values of $T_{\rm vis}$ for CO$_{2}$ and H$_{2}$O ice particles in this study, respectively.
The red star shows a reported value of $T_{\rm vis,\, H_{2}O}$ from numerical calculations by \citet{Gundlach+2015}.
The dashed lines are the (empirical) fitting formulae of size-dependent $T_{\rm vis,\, CO_{2}}$ and $T_{\rm vis,\, H_{2}O}$.
}
\label{fig6}
\end{figure}

Then we also apply the empirical relation to the size-dependent relaxation time of CO$_{2}$ ice particles.
From Relation (\ref{T_vis_CO2}), we found that the following relation,
\begin{equation}
T_{\rm vis,\, CO_{2}} = 1 \times 10^{-11}\ {\left( \frac{R_{1}}{1\ \upmu{\rm m}} \right)}^{1.11}\ {\rm s},
\label{T_vis_R1_CO2}
\end{equation}
is consistent with the experimental results for CO$_{2}$ ice particles with $R_{1} = 60\ \upmu{\rm m}$.
Figure \ref{fig6} shows the dependence of $T_{\rm vis}$ on $R_{1}$ for both CO$_{2}$ and H$_{2}$O ice particles.

\subsection{Size dependence of threshold velocity for sticking}

Here we calculate $V_{\rm stick}$ for (sub)$\upmu{\rm m}$-sized CO$_{2}$ ice particles by using the size-dependent relaxation time derived in Section \ref{sec.5.3}.
We set $\gamma_{\rm SV} = 80\ {\rm mJ}\ {\rm m}^{-2}$, $E = 10.7\ {\rm GPa}$, $\nu = 0.26$, and $\rho = 1650\ {\rm kg}\ {\rm m}^{-3}$ (see Section \ref{sec.3}).

\begin{figure*}
\plottwo{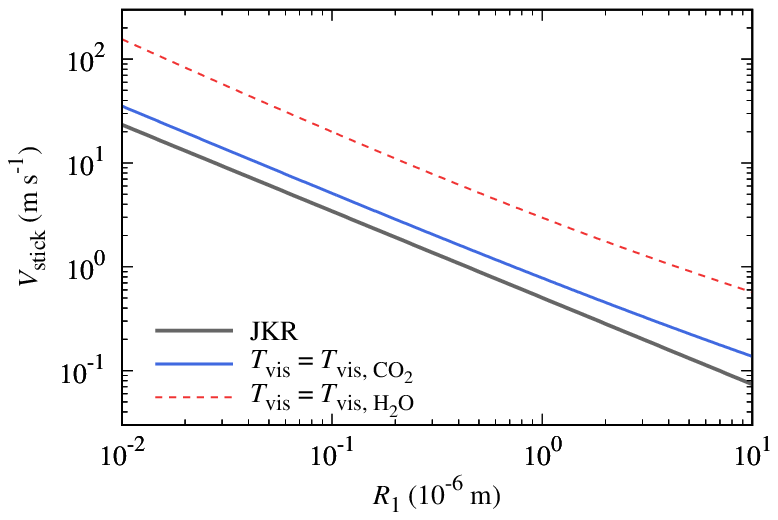}{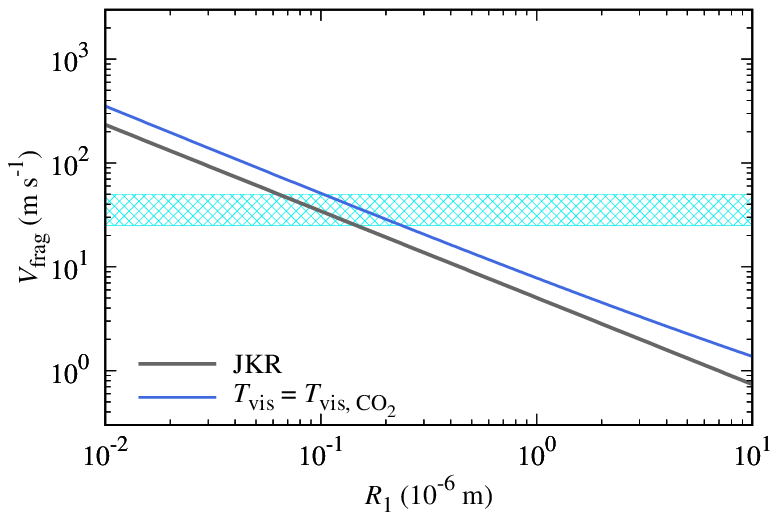}
\caption{Dependence of the threshold velocity for sticking (for collisions between monomer grains), $V_{\rm stick}$, and the critical velocity for collisional growth/fragmentation (for inter-aggregate collisions), $V_{\rm frag}$, as functions of the particle radius of monomer grains, $R_{1}$.
The left panel shows $V_{\rm stick}$ for different assumptions for the size-dependent relaxation time.
The red dashed line is $V_{\rm stick}$ for $T_{\rm vis} = T_{\rm vis,\, H_{2}O}$.
The blue solid line is $V_{\rm stick}$ for $T_{\rm vis} = T_{\rm vis,\, CO_{2}}$, and this is the standard model in this study.
The grey solid line shows $V_{\rm stick,\, JKR}$ as a lower limit of $V_{\rm stick}$.
The right panel shows $V_{\rm frag}$ for different assumptions for the size-dependent relaxation time.
The blue solid line is $V_{\rm frag}$ for $T_{\rm vis} = T_{\rm vis,\, CO_{2}}$.
The grey solid line also shows $V_{\rm frag,\, JKR}$ as a lower limit of $V_{\rm frag}$.
In our estimates, $V_{\rm frag}$ is given by $V_{\rm frag} = 10 V_{\rm stick}$ (Equation \ref{eq.vfrag}).
}
\label{fig7}
\end{figure*}

The left panel of Figure \ref{fig7} shows $V_{\rm stick}$ as a functin of $R_{1}$.
We also consider the dependence of $T_{\rm vis}$ on $V_{\rm stick}$.
The grey line represents the case of perfectly elastic contact model, i.e., $V_{\rm stick} = V_{\rm stick,\, JKR}$.
The blue line represents the standard model, i.e., $T_{\rm vis} = T_{\rm vis,\, CO_{2}}$ (Equation \ref{T_vis_R1_CO2}).
We can find that the difference between these two models is within a factor of a few in $V_{\rm stick}$, and we can (roughly) evaluate $V_{\rm stick}$ by using JKR theory, which is widely used in previous studies \citep[e.g.,][]{Dominik+1997,Wada+2007}.

In contrast, the viscoelastic dissipation effects play a great role when $T_{\rm vis}$ is several times higher than that we assumed for CO$_{2}$ ice particles.
The red dashed line is the threshold velocity for sticking, $V_{\rm stick}$, for the case when $T_{\rm vis} = T_{\rm vis,\, H_{2}O}$ (Equation \ref{T_vis_R1_H2O}).
As $V_{\rm stick}$ is an order of magnitude higher than $V_{\rm stick,\, JKR}$ when we use $T_{\rm vis} = T_{\rm vis,\, H_{2}O}$, we can imagine that the large difference of $V_{\rm stick}$ between CO$_{2}$ and H$_{2}$O ice particles \citep{Musiolik+2016a,Musiolik+2016b} mainly originate from the large difference of $T_{\rm vis}$ between two materials.

\citet{Pinilla+2017} and \citet{Okuzumi+2019} mentioned that the low value of $V_{\rm stick}$ for CO$_{2}$ ice particles is due to the small surface free energy of {\it apolar} CO$_{2}$ ice.
However, the literature value of $\gamma_{\rm SV} = 80\ {\rm mJ}\ {\rm m}^{-2}$ \citep{Wood1999} is comparable to that of H$_{2}$O ice, although future direct measurements of the surface free energy of CO$_{2}$ ice is essential.
In addition, the values of elastic properties, $E$ and $\nu$, are also similar between two materials.
Therefore, we proposed that the large difference in $V_{\rm stick}$ between CO$_{2}$ and H$_{2}$O ice particles is thought to originate from the large difference in $T_{\rm vis}$.

\section{Critical Velocity for Collisional Fragmentation of Aggregates}
\label{sec.6}

Here we discuss the critical velocity for collisional growth/fragmentation, $V_{\rm frag}$, of dust aggregates composed of $\upmu{\rm m}$-sized monomer grains.
The right panel of Figure \ref{fig7} shows the dependence of $V_{\rm frag}$ on $R_{1}$.
Here $R_{1}$ is the radius of monomer grains.
The cyan hatched region indicates the maximum collision velocity of dust aggregates in circumstellar disks with weak turbulence, i.e., $25\ {\rm m}\ {\rm s}^{-1} \lesssim V_{\rm col,\, max} \lesssim 50\ {\rm m}\ {\rm s}^{-1}$ \citep[e.g.,][]{Adachi+1976,Blum+2008,Wada+2013}.

It is empirically known that $V_{\rm frag}$ is an order of magnitude larger than $V_{\rm stick}$ and is almost independent of the number of constituent monomer grains \citep{Dominik+1997,Wada+2009,Wada+2013}.
Here we briefly explain the basic findings from numerical simulations of collisions of dust aggregates.
Based on JKR theory, the amount of energy dissipated in a bouncing collision, $E_{\rm stick,\, JKR}$, is given by \citep{Thornton+1998,Wada+2007}
\begin{eqnarray}
E_{\rm stick,\, JKR} & = & \frac{m^{*}}{2} {V_{\rm stick,\, JKR}}^{2} \nonumber \\
                     & = & 0.9355 F_{\rm crit} \delta_{\rm crit},
\end{eqnarray}
where $F_{\rm crit} = 3 \pi \gamma R / 2$ is the maximum force needed to separate two contact particles and $\delta_{\rm crit} = {\left( 9 / 16 \right)}^{1/3} \delta_{\rm eq}$ is the critical pulling length between the particles in contact.
The energy necessary to break completely a contact in the equilibrium position, $E_{\rm break,\, JKR}$, is slightly larger than $E_{\rm stick,\, JKR}$ \citep[e.g.,][]{Wada+2007}:
\begin{equation}
E_{\rm break,\, JKR} = 1.54 F_{\rm crit} \delta_{\rm crit},
\end{equation}
and we usually use $E_{\rm break,\, JKR}$ to interpret collision outcomes of dust aggregates.

According to \citet{Wada+2013}, the critical velocity for collisional growth/fragmentation of dust aggregate of perfectly elastic monomer grains, $V_{\rm frag,\, JKR}$, is empirically given by
\begin{eqnarray}
V_{\rm frag,\, JKR} & = & C \sqrt{\frac{E_{\rm break,\, JKR}}{m_{1}}} \nonumber \\
                    & = & 0.64 C \cdot V_{\rm stick,\, JKR},
\end{eqnarray}
where $C$ is a dimensionless constant: $C \simeq 15$ for equal-sized collisions and $C \simeq 20$ for different-sized collisions \citep{Wada+2009,Wada+2013}.\footnote{
\citet{Wada+2009,Wada+2013} numerically revealed that the value of $C$ hardly depends on the size of aggregates when the number of constituent monomer grains is in the range between $10^{3}$ and $10^{6}$.
We note, however, that the detailed reason why $C$ hardly depends on the size of aggregates is still unclear.
}
Therefore, the scaling relation between $V_{\rm frag,\, JKR}$ and $V_{\rm stick,\, JKR}$ is approximately given by $V_{\rm frag,\, JKR} = 10 V_{\rm stick,\, JKR}$.
Although it is not clear that whether this relation between $V_{\rm frag}$ and $V_{\rm stick}$ is appliable for dust aggregates of viscoelastic monomer grains \citep[e.g.,][]{gunkelmann2016}, we apply the following assumption to evaluate the value of $V_{\rm frag}$:
\begin{equation}
V_{\rm frag} = 10 V_{\rm stick}.
\label{eq.vfrag}
\end{equation}

The right panel of Figure \ref{fig7} suggests that $V_{\rm frag} > V_{\rm col,\, max}$ when the radius of monomer grains is $R_{1} \ll 0.1\ \upmu{\rm m}$, and $V_{\rm frag} < V_{\rm col,\, max}$ for the case of $R_{1} \gg 0.1\ \upmu{\rm m}$.
In the context of dust growth in circumstellar disks, we usually assumed $R_{1} = 0.1\ \upmu{\rm m}$ in numerical calculations \citep[e.g.,][]{Okuzumi+2012,Krijt+2015,Homma+2018}.
This assumption is at least consistent with the grain size in the surrounding envelope of proto-stellar objects inferred from near-infrared polarimetry \citep[e.g.,][]{Murakawa+2008} and the size distribution of interstellar dust grains \citep[e.g.,][]{Mathis+1977,Weingartner+2001}.
For $R_{1} = 0.1\ \upmu{\rm m}$, dust aggregates of CO$_{2}$ ice monomer grains can stick together without catastrophic fragmentation when the strength of turbulence is weak.
In this case, the maximum size of dust aggregates is controlled not by fragmentation but by radial drift \citep[e.g.,][]{Okuzumi+2012,Drazkowska+2017}, although bouncing and/or erosive collisions between particles with a high mass ratio might prevent dust aggregates from growing into larger aggregates \citep[e.g.,][]{Zsom+2010,Krijt+2015}.
In Section \ref{sec.7.2}, we discuss the possible mechanisms for altering the size of monomer grains.

\section{Discussion}
\label{sec.7}

\subsection{Morphology of ice particles used in experiments}
\label{sec.7.1}

We consider that CO$_{2}$ ice particles used in \citet{Musiolik+2016a} may be monolithic (see Section \ref{sec.5.1}).
This is because the critical velocity for collisional growth/fragmentation, $V_{\rm frag}$, is several times higher than $0.04\ {\rm m}\ {\rm s}^{-1}$ when we assume that $60\ \upmu{\rm m}$-sized CO$_{2}$ ice particles are dust aggregates.
If the aggregate radius is $60\ \upmu{\rm m}$, then the radius of monomer grains should be smaller than the half of the aggregate radius, i.e., $R_{1} \lesssim 30\ \upmu{\rm m}$.
In this case, $V_{\rm frag,\, JKR}$ of dust aggregates composed of CO$_{2}$ ice monomer grains with $R_{1} \lesssim 30\ \upmu{\rm m}$ is
\begin{eqnarray}
V_{\rm frag,\, JKR} & = & 10 V_{\rm stick,\, JKR} \nonumber \\
                     & = & 2.96 \times 10^{-1}\ {\left( \frac{R_{1}}{30\ \upmu{\rm m}} \right)}^{- 5/6}\ {\rm m}\ {\rm s}^{-1}.
\end{eqnarray}
Moreover, this large value of $V_{\rm frag,\, JKR}$ gives the minimum estimate of $V_{\rm frag}$.
Therefore, the estimated $V_{\rm frag}$ is an order of magnitude higher than the threshold velocity reported by \citet{Musiolik+2016a}.

We also note that the experimental setup of \citet{Musiolik+2016b} is identical to that of \citet{Musiolik+2016a}.
The particle radius of CO$_{2}$ and H$_{2}$O ices are very similar: $R_{1} = 60\ \upmu{\rm m}$ and $90\ \upmu{\rm m}$, respectively.
Therefore, we conclude that both CO$_{2}$ and H$_{2}$O ice particles used in \citet{Musiolik+2016a,Musiolik+2016b} are not aggregates but monolithic grains.

\subsection{Size of monomer grains}
\label{sec.7.2}

The size of monomer grains is often taken to $0.1\ \upmu{\rm m}$ \citep[e.g.,][]{Okuzumi+2012}; however, it is unclear to what extent using a single and constant monomer size is appropriate.
Here we discuss several possible scenarios that can alter the size of monomer grains in circumstellar disks.

\citet{Ros+2013} proposed that condensation of H$_{2}$O vapor near the H$_{2}$O snow line might be a dominant particle growth mechanism when dust growth is prevented by bouncing and/or fragmentation.
If condensation of H$_{2}$O vapor controls the size of monomer grains, the physics of heterogeneous nucleation may play a crucial role.
Laboratory experiments on heterogeneous nucleation by \citet{Iraci+2010} revealed that the formation of a H$_{2}$O ice layer on a bare silicate surface requires a substantially high H$_{2}$O vapor pressure.
Then \citet{Ros+2019} showed that H$_{2}$O vapor may be deposited predominantly on already ice-covered particles and these icy particles can grow into ${\rm cm}$-sized huge monomer grains near the H$_{2}$O snow line.
In this scenario, ${\rm cm}$-sized huge monomer grains cannot agglomerate because $V_{\rm stick}$ for ${\rm cm}$-sized huge monomer grains is too low even if they are covered by a H$_{2}$O ice mantle.
Then icy planetesimals might be formed through gravitational collapse of clumps of ${\rm cm}$-sized icy grains \citep[e.g.,][]{Johansen+2007,Bai+2010}.

This selective condensation process may be important not only near the H$_{2}$O snow line but also near the CO$_{2}$ snow line.
We note, however, that the formation process of the first CO$_{2}$ ice layer may different from that of the H$_{2}$O ice layer.
As CO$_{2}$ ice might be formed via chemical reaction of CO and OH on grain surfaces \citep[e.g.,][]{Bosman+2018,Krijt+2020}, the size of monomer grains covered by a CO$_{2}$ ice mantle would be similar to that covered by a H$_{2}$O ice mantle near the CO$_{2}$ snow line.

Another possible mechanism for changing the size of monomer grains is evaporation and following recondensation of dust particles via flash-heating events \citep[e.g.,][]{Miura+2010,Arakawa+2016}.
The flash-heating events in the early solar nebula are thought to be the plausible formation mechanisms of chondrules contained within chondrites \citep[e.g.,][]{Arakawa+2019}.
Recently, \citet{Fujiya+2019} revealed that at least some chondrite parent bodies were formed beyond the CO$_{2}$ snow line, based on C-isotope measurements on carbonate minerals in carbonaceous chondrites.
Then the flash-heating events might occur not only in the inner region of the solar nebula but also outside the CO$_{2}$ snow line, and the following recondensation process would determine the size of monomer grains in the early solar nebula.

Based on the combination of dust evolution calculations and synthetic polarimetric observations of a circumstellar disk around a young star HL Tau, \citet{Okuzumi+2019} revealed that the plausible value of $V_{\rm frag}$ is lower than $1\ {\rm m}\ {\rm s}^{-1}$ both inside the H$_{2}$O snow line and outside the CO$_{2}$ snow line, to explain the small dust scale height \citep{Pinte+2016} and the observed aggregate radius of $\simeq 100\ \upmu{\rm m}$ \citep[e.g.,][]{Kataoka+2017,Stephens+2017} simultaneously.
This suggests that the size of monomer grains in the disk around HL Tau might be $R_{1} \gtrsim 10\ \upmu{\rm m}$ (see right panel of Figure \ref{fig7}), and some mechanisms for altering the size of monomer grains from that of interstellar dust grains are required.

\subsection{Impact of dust growth on the gas-phase abundance of carbon monoxide in circumstellar disks}

Understanding the astrochemistry of CO in circumstellar disks is of great importance in the context of star and planet formation.
This is because emission from gas-phase CO and its isotopologues is widely used to study the structures of circumstellar disks, such as the disk radius \citep[e.g.,][]{Ansdell+2018}, the disk mass \citep[e.g.,][]{Ansdell+2016}, the temperature profile \citep[e.g.,][]{Dullemond+2020}, and the presence of giant planets \citep[e.g.,][]{Pinte+2019}.

It is known that the abundance of CO relative to hydrogen in circumstellar disks decreases by up to factors of 10--100 from its interstellar medium value \citep[e.g.,][]{Bergner+2020,Zhang+2020}, and there are a large number of papers which studied chemical processing of CO as the origin of its depletion \citep[e.g.,][]{Aikawa+1996,Bergin+2014,Furuya+2014,Bosman+2018}.
Physical sequestration of CO ice can also contribute the depletion of gas-phase CO \citep[e.g.,][]{Kama+2016,xu2017,Krijt+2018}.
As the vertical settling of large and icy dust aggregates called ``pebbles'' is the key mechanism of sequestration, dust growth and ensuing radial drift \citep[e.g.,][]{Zhang+2020b} are directly associated with the gas-phase abundance of CO.
If icy monomer grains are indeed submicron-sized spheres, our results suggest that collisional growth is unlikely to be hindered by fragmentation in the cold outer region of circumstellar disks.

\section{Summary}
\label{sec.8}

We have investigated the reason for the low threshold velocity for sticking of CO$_{2}$ ice particles compared to that of H$_{2}$O ice particles.
Using the viscoelastic contact model \citep{Krijt+2013}, we succeeded in reproducing the experimental results of collisions of CO$_{2}$ and H$_{2}$O ice particles \citep{Musiolik+2016a,Musiolik+2016b}.
Our findings are summarized as follows.

\begin{enumerate}
\item{
For collisons between two viscoelastic spheres, we found that there are three types of collision outcomes, namely, sticking collisions, bouncing collisions, and double collisions.
We defined the threshold velocity for sticking, $V_{\rm stick}$, as the transition velocity from double collisions to bouncing collisions (see Figures \ref{fig2} and \ref{fig3}).
}
\item{
In the viscoelastic contact model, the relaxation time, $T_{\rm vis}$, is the key parameter to describe the strength of viscoelastic effects \citep{Krijt+2013}.
We found that the relaxation time of CO$_{2}$ ice particles with the particle radius of $R_{1} = 60\ \upmu{\rm m}$ is in the range of $8.5 \times 10^{-11}\ {\rm s} \leq T_{\rm vis} \leq 1.97 \times 10^{-9}\ {\rm s}$, and $V_{\rm stick}$ of CO$_{2}$ ice particles is not so different from that predicted from JKR theory for perfectly elastic spheres (see Figure \ref{fig4}).
}
\item{
In contrast, we found that $V_{\rm stick}$ of H$_{2}$O ice particles is an order of magnitude higher than that predicted from JKR theory (see Figure \ref{fig5}).
The relaxation time of H$_{2}$O ice particles with the particle radius of $R_{1} = 90\ \upmu{\rm m}$ should be in the range of $5.3 \times 10^{-9}\ {\rm s} \leq T_{\rm vis} \leq 2.37 \times 10^{-8}\ {\rm s}$, and this value of $T_{\rm vis}$ is an order of magnitude higher than that for CO$_{2}$ ice particles.
This relaxation time for H$_{2}$O ice particles obtained from our numerical results is consistent with the result of \citet{Gundlach+2015} when we use the empirical relation between $T_{\rm vis}$ and $R_{1}$ (see Figure \ref{fig6}).
}
\item{
Therefore, we concluded that the large difference in stickiness between H$_{2}$O and CO$_{2}$ ice particles would mainly originate from the difference in the strength of viscoelastic effects.
}
\item{
We also evaluated the critical velocity for collisional growth/fragmentation, $V_{\rm frag}$, of dust aggregates composed of $\upmu{\rm m}$-sized CO$_{2}$ ice particles.
Assuming that $V_{\rm frag}$ is approximately given by $V_{\rm frag} = 10 V_{\rm stick}$ and the radius of monomer grains is $R_{1} = 0.1\ \upmu{\rm m}$, we found that the maximum size of dust aggregates would be controlled not by fragmentation but by radial drift even outside the CO$_{2}$ snow line (see Figure \ref{fig7}).
}
\end{enumerate}

More broadly, our results highlight the importance of additional energy dissipation channels during collisions of dust particles.
Thus future studies on the (viscoelastic) material properties of ices, including  H$_{2}$O, CO$_{2}$, CO, CH$_{4}$, CH$_{3}$OH, and NH$_{3}$, are of great importance to understand the physics and chemistry in circumstellar disks.
We also need to study the interplay between dust growth and chemical evolution in circumstellar disks.

\acknowledgments

We would like to thank Stephen E.\ Wood for providing information about the surface energy of CO$_{2}$ ice.
S.A.\ is very thankful to Kenji Furuya for fruitful discussions.
S.A.\ is supported by JSPS KAKENHI Grant No.\ JP20J00598.

\appendix
\restartappendixnumbering

\section{Dependence of apparent surface energy on crack speed}
\label{App.A}

\begin{figure*}
\plottwo{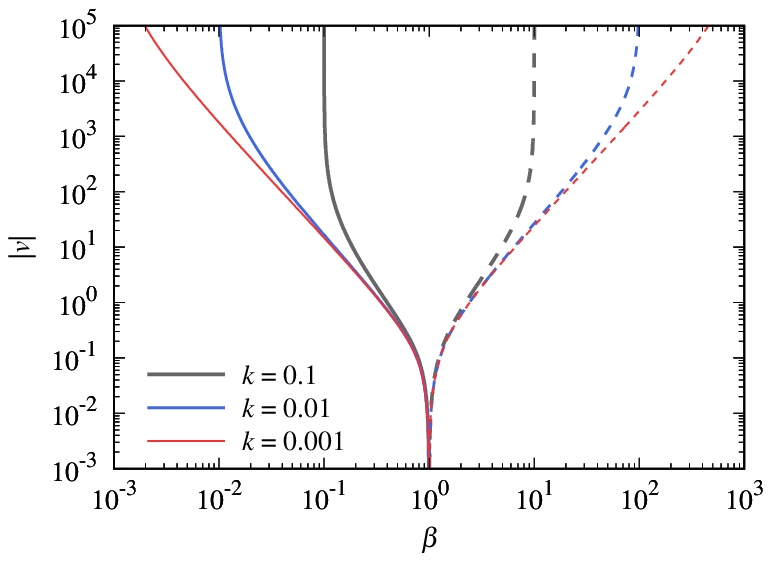}{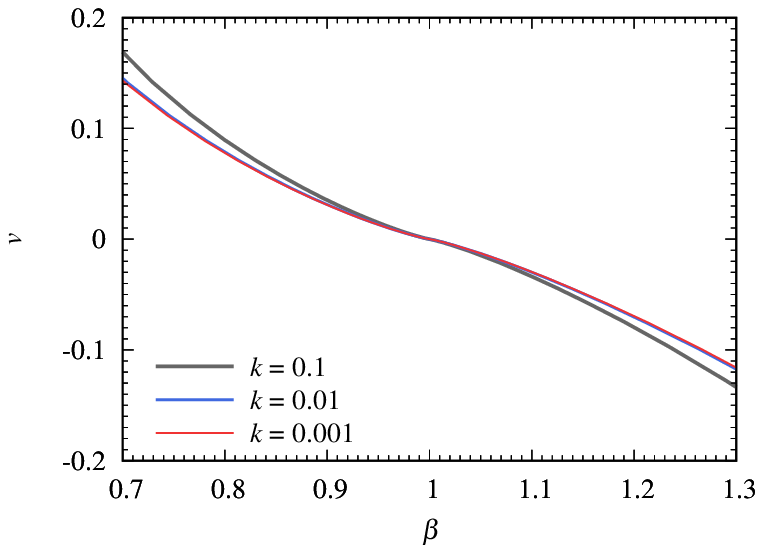}
\caption{
Dependence of the normalized crack velocity, $v$, on the normalized apparent surface energy, $\beta$.
The left panel shows $v = v {\left( k, \beta \right)}$ within the range of $10^{-3} < \beta < 10^{3}$, and the right panel shows $v = v {\left( k, \beta \right)}$ near $\beta = 1$.
We note that $v$ is positive (i.e., $\dot{a} > 0$) when $\beta < 1$ and $v$ is negative when $\beta > 1$, and $v = 0$ at $\beta = 1$ \citep[see][]{Greenwood2004}.
}
\label{fig1}
\end{figure*}

\citet{Greenwood2004} derived the apparent surface energy which depends on the crack velocity using the Maugis--Dugdale model of the surface force law \citep{Dugdale1960,Maugis1992}.
The normalized apparent surface energy and crack velocity, $\beta$ and $v$, are given as functions of $k$ and $\alpha$, where $\alpha$ is the non-dimensional transit time \citep{Greenwood2004}.
For the opening crack, $\beta$ and $v$ are given by
\begin{eqnarray}
\beta & = & \frac{1}{I_{1} {\left( k, \alpha \right)}}, \\
v & = & - \frac{\pi}{4} \frac{1}{\alpha {I_{1} {\left( k, \alpha \right)}}},
\end{eqnarray}
and for the closing crack,
\begin{eqnarray}
\beta & = & \frac{{\left[ I_{3} {\left( k, \alpha \right)} \right]}^{2}}{I_{2} {\left( k, \alpha \right)}}, \\
v & = & \frac{\pi}{4} \frac{1}{\alpha {I_{2} {\left( k, \alpha \right)}}}.
\end{eqnarray}
Here $I_{1}$, $I_{2}$, and $I_{3}$ are given by
\begin{eqnarray}
I_{1} {\left( k, \alpha \right)} & = & k + {\left( 1 - k \right)} J_{1} {\left( \alpha \right)}, \\
I_{2} {\left( k, \alpha \right)} & = & k + {\left( 1 - k \right)} J_{2} {\left( \alpha \right)}, \\
I_{3} {\left( k, \alpha \right)} & = & 1 - {\left( 1 - k \right)} J_{3} {\left( \alpha \right)},
\end{eqnarray}
and $J_{1}$, $J_{2}$, and $J_{3}$ are functions of $\alpha$:
\begin{eqnarray}
J_{1} {\left( \alpha \right)} & = & \frac{1}{2} \alpha \int_{0}^{1} {\rm d}\xi\ \exp{\left[ - \alpha {\left( 1 - \xi \right)} \right]} {F {\left( \xi \right)}}, \\
J_{2} {\left( \alpha \right)} & = & \frac{1}{2} \alpha \int_{0}^{1} {\rm d}\xi\ \exp{\left[ - \alpha {\left( 1 - \xi \right)} \right]} {A {\left( \xi \right)}}, \\
J_{3} {\left( \alpha \right)} & = & \frac{1}{2} \int_{0}^{1} {\rm d}\xi\ \exp{\left[ - \alpha {\left( 1 - \xi \right)} \right]} \xi^{- 1/2}, \\
F {\left( \xi \right)} & = & 2 \sqrt{\xi} - {\left( 1 - \xi \right)} \log{\left| \frac{1 + \sqrt{\xi}}{1 - \sqrt{\xi}} \right|}, \\
A {\left( \xi \right)} & = & 2 \sqrt{\xi} + {\left( 1 - \xi \right)} \log{\left| \frac{1 + \sqrt{\xi}}{1 - \sqrt{\xi}} \right|}.
\end{eqnarray}
As both $\beta$ and $v$ are the functions of $\alpha$, we can regard $\alpha$ as an auxiliary variable.
Then we obtained $v = v {\left( k, \beta \right)}$ as shown in Figure \ref{fig1}.
We also note that Tables 1 and 2 of \citet{Greenwood2004} show the values of $v$ and $\beta$ as functions of $\alpha$, for the case of $k = 0.01$.
It should be noted that $v$ is defined within the range of $k < \beta < 1 / k$.
For the opening crack, $\beta \to 1/k$ and $v \to - \infty$ when $\alpha \to 0$, and for the closing crack, $\beta \to k$ and $v \to + \infty$ when $\alpha \to 0$.

\bibliography{sample63}{}
\bibliographystyle{aasjournal}

\end{document}